\documentclass[12pt]{iopart}

\usepackage{iopams}
\usepackage[dvips]{graphicx}

\begin{document}

\title{How to retrieve additional information from the multiplicity distributions}

\author{Grzegorz Wilk$^1$, Zbigniew W\l odarczyk$^2$}
\address{$^1$National Centre for Nuclear Research,  Department of Fundamental Research, Ho\.za 69, 00-681
        Warsaw, Poland} \ead{grzegorz.wilk@ncbj.gov.pl}
\address{$^2$Institute of Physics, Jan Kochanowski University,
\'Swi\c{e}tokrzyska 15, 25-406 Kielce, Poland}
\ead{zbigniew.wlodarczyk@ujk.kielce.pl}

\begin{abstract}
Multiplicity distributions $P(N)$ measured in multiparticle production processes are most frequently described by the Negative Binomial Distribution (NBD). However, with increasing collision energy some systematic discrepancies have become more and more apparent. They are usually attributed to the possible multi-source structure of the production process and described using a multi-NBD form of the multiplicity distribution. We investigate the possibility of keeping a single NBD but with its parameters depending on the multiplicity $N$. This is done by modifying the widely known clan model of particle production leading to the NBD form of $P(N)$. This is then confronted with the approach based on the so-called cascade-stochastic formalism which is based on different types of recurrence relations defining $P(N)$. We demonstrate that a combination of both approaches allows the retrieval of additional valuable information from the multiplicity distributions, namely the oscillatory behavior of the counting statistics apparently visible in the high energy data.
\end{abstract}


\submitto{\jpg}

\maketitle

\section{Introduction}
\label{intro}

Multiplicity distributions $P(N)$ measured in multiparticle production processes are a source of valuable information on the dynamics of these processes and are among the first observables measured and intensively studied in any multiparticle production experiment. Remarkably, essentially for all collision energies studied so far, ranging from tenths to thousands of GeV, the most commonly used form of $P(N)$ is the two-parameter negative binomial distribution (NBD) function (see reviews
\cite{PPP,DG,CS,DWDK,GU,Kittel,Kanki1,Kanki2}),
\begin{equation}
P(N;p,k) = \frac{\Gamma(N+k)}{\Gamma(k)\Gamma(N+1)}
p^N (1 - p)^k,
\label{NBD}
\end{equation}
where
\begin{equation}
p = p(m,k) = \frac{m}{m + k} \label{partem}
\end{equation}
is the probability of particle emission. The $N$ is the observed number of particles and  $m$ and $k$ are the two parameters of the NBD.  The operational definition tells us that $m$ is connected with the measured multiplicity, $m = \langle N\rangle$ (at least for $m = const$). The actual meaning of the parameters $m$ and $k$ depends on the particular dynamical description of the measured $P(N)$ resulting in the NBD. Its history begins almost a century ago \cite{Old}. It was then realized that the NBD emerges whenever one has to account for a fluctuating medium \cite{VP}, or for a system which is intrinsically nonextensive \cite{WWq}. It also emerges from the information theory (Shannon entropy) approach \cite{MaxEnt}. In multiparticle production phenomenology the NBD has been obtained in the bag model \cite{MGBM}, in the stochastic bootstrap model \cite{SBM,SBM1}, in numerous descriptions using branching and stochastic processes \cite{MP,SBW,TA}, in popular and frequently used clan model \cite{GU,CMa2,CMb2} (for its more theoretical justification see \cite{CMTh1,CMTh2}), or using a general form of the grand canonical partition function \cite{AM1,AM2,AM3}. It also describes in a natural way different kinds of multiparton interactions \cite{MPI,WDW,DN} and the production of some specific initial gluonic states in the first phase of the interaction process, the so called "glittering glasmas" \cite{GLMc}. For the purpose of this work we shall concentrate on the not so widely known  cascade-stochastic model \cite{CSF,CSF1}. The multiplicity distribution (MD) can be defined in different ways. For our further consideration the most suitable are the recurrence relations between some selected distributions. The simplest one is the relation between adjacent distributions, $P(N)$ and $P(N+1)$, only. This corresponds to the assumption of a connection existing only between the production of $N$ and $N + 1$ particles:
\begin{equation}
(N+1)P(N+1) = g(N)P(N). \label{RR}
\end{equation}
The type of MD described by it is determined by the function $g(N)$. The simplest nontrivial choice is given by a linear relation to multiplicity $N$,
\begin{equation}
g(N) = \alpha + \beta N \label{g(N)}.
\end{equation}
 This covers, for example, the Poissonian distribution (for $ \alpha =  \langle N\rangle$ and $\beta =0$), binomial distribution (for $\alpha = \langle N\rangle k/(k -\langle N\rangle)$ and $\beta = - \alpha/k$) and NBD (for $\alpha = \langle N\rangle k/(k +\langle N\rangle)$ and $ \beta = \alpha/k$). Usually, when searching for the best MD to fit the experimental data, one modifies accordingly the function $g(N)$ (for example, by introducing higher order terms \cite{HC} or by using its more involved form\emph{\emph{}}s \cite{MF,Z1}).

The more general form of recurrence relation connects the multiplicity $N+1$ with all smaller multiplicities and has the form \cite{ST}:
\begin{equation}
(N +1)P(N + 1) = \langle N\rangle \sum_{j=0}^NC_j P(N-j).
\label{recPN}
\end{equation}
Coefficients $C_j$ now define the corresponding MD $P(N)$. For the Poisson distribution $C_0 = 1$ and all $C_{j>0} =0$. Such a form of the recurrence relation occurs in a natural way in the so called \emph{cascade-stochastic processes} \cite{ST,CMC}. In \cite{CSF,CSF1} it was successfully applied to multiparticle phenomenology.  As will be shown below, this recurrence relation allows some additional information to be retrieved from the experimentally measured multiplicity distributions.

\begin{figure}
\begin{center}
\resizebox{0.6\textwidth}{!}{%
  \includegraphics{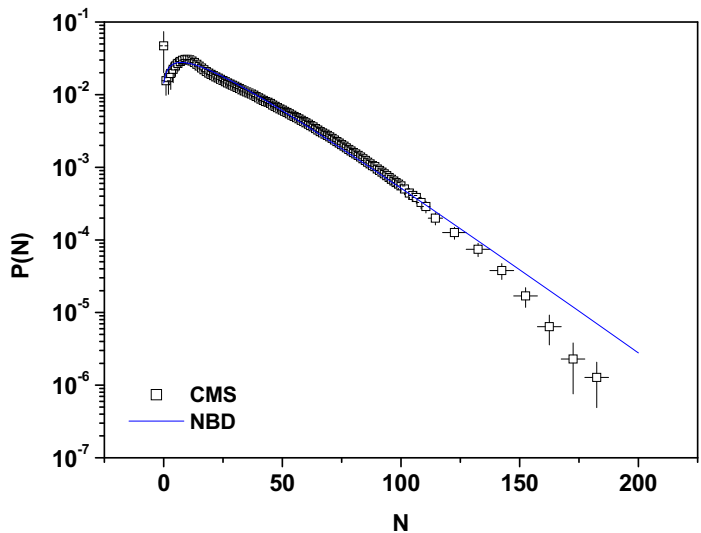}
} \vspace{-5mm}\caption{(Color online) Charged hadron multiplicity distributions for  $|\eta| < 2$  at $\sqrt{s} =7$ TeV, as given by the CMS experiment \cite{CMS} (points), compared with the NBD for parameters $m = 25.5$ and $k = 1.45$ (solid line).}
\label{Fig_NBD}
\end{center}
\end{figure}

\begin{figure}
\begin{center}
\resizebox{0.6\textwidth}{!}{%
  \includegraphics{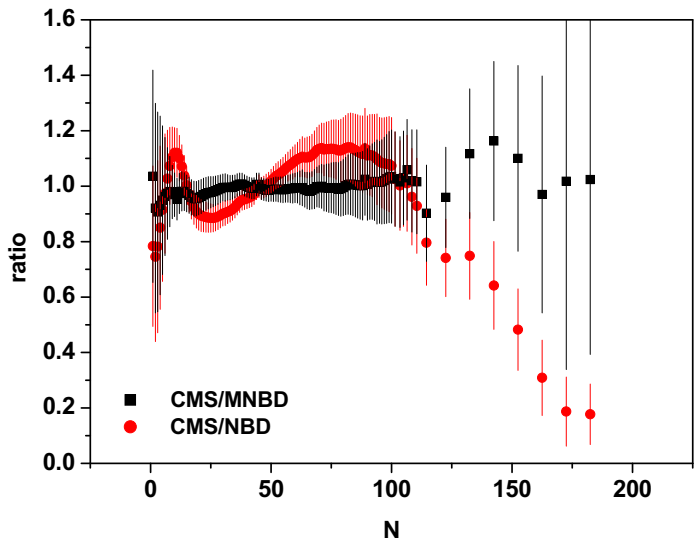}
} \vspace{-5mm}\caption{(Color online) Multiplicity dependence of the ratio $R = P_{CMS}(N)/P_{NBD}(N)$ for the data shown in Fig. \ref{Fig_NBD} (red circles) and of the corrected ratio $R$ (black squares) obtained from the MNBD discussed below using the parametrization (\ref{Non}).} \label{Fig_Modified}
\end{center}
\end{figure}

Notwithstanding the popularity of the NBD, closer inspection shows some systematic deviations of fits based on the NBD from data which become more pronounced with increasing energy. Starting from the SPS and Tevatron energies \cite{PPP}, where a kind of shoulder structure appeared in the multiplicity distribution, the recent ALICE \cite{ALICE,ALICEa,ALICE1} and CMS \cite{CMS} data at LHC show that a single NBD formula works, at best, only in some limited range of multiplicities. This can be seen in Fig. \ref{Fig_NBD} showing results from CMS where for large multiplicities the experimental points are below the best single NBD fit. Even more dramatic differences are visible in plots of the ratio $R = P_{CMS}(N)/P_{NBD}(N)$, see Fig. \ref{Fig_Modified} (red circles). In addition to the falling tail one can also see structure for smaller multiplicities.

The appearance of such distinctive substructures in the multiplicity distributions at higher energies is usually attributed to the weighted (incoherent) superposition of more than one source of particle production, each producing particles according to some MD, with the final MD being their sum (cf., for
example, \cite{GU,MPI,WDW,DN,FFWW,PG} or \cite{PPP,DG} and references therein). In fact, the observed substructure in the SPS data at $900$ GeV  and in the Tevatron data at $1.8$ TeV  can be explained by a weighted superposition of two NBD functions \cite{GU}, whereas the recent LHC data are fitted using two  \cite{ALICE1,PG}, three \cite{Z}, or even more \cite{Z1} NBDs.

In this paper we undertake a different approach. First we investigate what kind of changes in the structure of the original MD, taken in the NBD form, are necessary in order to describe the same data by a single NBD with accordingly modified parameters $m$ and $k$. To be specific, in Section \ref{sec-MNBD} we formulate a modified negative binomial distribution (MNBD) in such a way as to reach agreement with data for the ratio $R = P_{CMS}(N)/P_{MNBD}(N)$.  Our result is presented in Fig. \ref{Fig_Modified} as the black squares. We proceed further and in Section \ref{CS} we investigate the possibility of retrieving some additional information from the measured $P(N)$. To this end we use the recurrence relation given by Eq. (\ref{recPN}) but apply it in a different way than is usually done. Namely, we use the experimentally measured $P(N)$ as input to calculate the corresponding coefficients $C_j$. The result obtained is remarkable. It turns out that the coefficients $C_j$ corresponding to data on $P(N)$ show intriguing oscillatory behavior, so far not disclosed and not discussed. It turns out that, so far, this behavior has not been obtained either from the usual NBD or from any simple combinations of NBDs used to fit these data (although, in principle, oscillations are possible in multi-component NBD scenarios). On the other hand, it arises when using the modified form of NBD obtained from the above mentioned MNBD used to fit data.

Our summary and conclusions are presented in Section \ref{SaC}.

\section{Modified Negative Binomial Distribution - MNBD}
\label{sec-MNBD}

To describe data using only a single multiplicity distribution, in our case a distribution based on the NBD, we allow the parameter $m$ to depend on the multiplicity $N$. It turns out that to get a flat distribution of the ratio $P_{data}(N)/P_{fit}(N)$ for all measured values of $N$ one needs a peculiar non-monotonic dependence of $m$ on $N$ given by\footnote{Such a change means that we preserve the overall form of the NBD given by Eq. (\ref{NBD}) because only the probability of particle emission $p$ is affected, cf. Eq. (\ref{partem}). Changes in the parameter $k$ would result in changing the form of the original NBD.},
\begin{equation}
m = m(N) = c\exp(a|N-b|), \label{NcN1}
\end{equation}
where $a$, $b$ and $c$ are parameters. This corresponds to a rather complicated, nonlinear and non monotonic form of $g(N)$ in  Eq.(\ref{RR}). Such a parametrization, with $a = 0.0455$, $b = 11$ and $c = 20.252$, improves the agreement with data. More careful examination shows that introducing additionally a small nonlinearity in $g(N)$, for example taking
\begin{equation}
m = c\exp\left[ a_1 | N - b| + a_2( N - b)^4\right], \label{Non}
\end{equation}
with $a_1 = 0.044$, $b = 11$, $ c = 20.252$ and $a_2 = 1.04\cdot 10^{-9}$, improves further the agreement with data as can be seen in Fig. \ref{Fig_Modified}.

\begin{figure}[t]
\begin{center}
\resizebox{0.6\textwidth}{!}{%
  \includegraphics{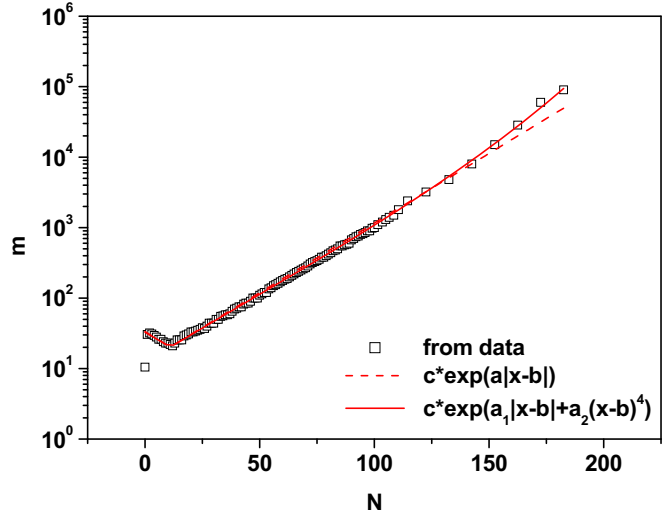}
} \vspace{-5mm}\caption{(Color online) The solid line corresponds to the dependence given by Eq.(\ref{Non}) and the dashed by Eq. (\ref{NcN1}).} \label{Fig_Modification}
\end{center}
\end{figure}

Both proposed dependencies of $m$ on $N$ (cf. Eqs. (\ref{NcN1}) and (\ref{Non})) shown in Fig. \ref{Fig_Modification} are valid for $N < N_{max}$, with some maximal cut-off due to the normalization of $P(N)$. In both cases the non-monotonic form of the proposed modification is clearly visible. It is located in the region of small multiplicities $N$. After it $m(N)$ grows steadily.  As already mentioned, this is the price to keep a single MD and to obtain the desired flat ratio shown in Fig. \ref{Fig_Modified} over the whole measured region of multiplicities $N$. At this moment we cannot offer any plausible interpretation of such behavior\footnote{Note that the spout-like form of the modification used here is just the simplest possible choice of parametrization that brings agreement with data.}.

\begin{figure}[t]
\begin{center}
\resizebox{0.6\textwidth}{!}{%
  \includegraphics{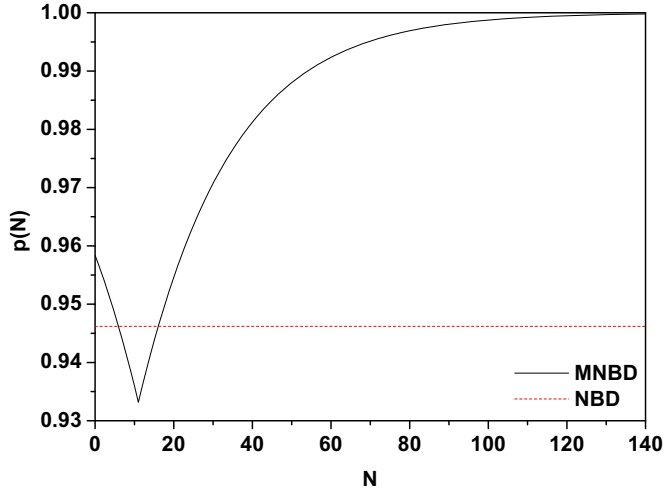}
} \vspace{-5mm}\caption{(Color online) Multiplicity dependence of the probability of particle emission, Eq. (\ref{partem}), for NBD and MNBD scenarios.} \label{Fig_PPE}
\end{center}
\end{figure}

Finally, note that the probability of particle emission, $p$, which is constant in the standard NBD (cf. Eq. (\ref{partem})),  in the MNBD depends on the multiplicity $N$ in the way presented in Fig. \ref{Fig_PPE}. The corresponding odds ratio in the MNBD is equal to $p(N)/[1 - p(N)] = m(N)/k$, i.e., the probability of emission ratio, $p(N)$, is given by the logistic function,
\begin{equation}
p(N) = \frac{1}{1 + \frac{k}{c}\exp ( - a |N - b|)}. \label{LF}
\end{equation}
The non-monoticity mentioned before is very pronounced in the $p(N)$ shown in Fig. \ref{Fig_PPE}.
 This fact will be crucial in our further discussion.

The other remark worth remembering is that in the production processes the important variable is the energy $W$ allowed for the production of particles\footnote{The fraction $W/\sqrt{s}$ is known as the \emph{inelasticity} and was widely used some time ago, cf. \cite{inel,inel1}. Recently it was related to the constituent quark picture under the name \emph{effective energy} \cite{EE1,EE2}.}. Assuming now that multiplicity $N$ depends logarithmically on $W$ \cite{PPP},
\begin{equation}
N = n_0 + n_1 \ln W \label{W}
\end{equation}
we can express $m(N)$ from Eq. ({\ref{NcN1}) as
\begin{equation}
m = c\left(\frac{W}{W_0}\right)^{\gamma},  \label{WW}
\end{equation}
where $\gamma = a n_1 sign\left(W - W_0\right)$ and $W_0$ is given by $b = n_0 + n_1 \ln W_0$.

\section{Oscillatory behavior of the counting statistics}
\label{CS}

In the experimental data the structure of the observed distributions $P(N)$ is not necessarily smooth. This fact is usually connected with the so called \emph{cascade-stochastic processes} \cite{ST,CMC} applied in multiparticle phenomenology in \cite{CSF,CSF1}, the description of which is based on the recurrence formula given by Eq. (\ref{recPN}). The multiparticle production has been visualized there as a two cascade process: the first being formation of groups of partons (generally speaking) which are then supposed to be converted into the observed hadrons. As a result the observed distributions consist of two connected probability cascades. Coefficients $C_j$ are then calculated from the respective rapidity distributions and finally one obtains the resulting $P(N)$.

However, we shall use Eq. (\ref{recPN}) in a different way. Namely, we ask what are the values of the coefficients $C_j$ corresponding to the experimentally observed particle distribution. Knowing $P(N)$ one can obtain the coefficients $C_j$ using the following recurrence formula:
\begin{equation}
\!\! \langle N\rangle C_j = (j + 1) \left[\frac{P(j+1)}{P(0)}\right] -
\langle N\rangle \sum^{j-1}_{i=0}C_i \left[ \frac{P(j-i)}{P(0)}\right]. \label{Cj}
\end{equation}
The result is striking. Namely, as can be seen in Fig. \ref{NBD_fa}, the coefficients $C_j$ obtained from the CMS data \cite{CMS} show distinct oscillatory behavior. This behavior can be fitted by a triangular wave, $C_j \propto (2/\pi) \arcsin[\sin ( 2\pi j/\omega)]$, damped exponentially by some exponential factor $\propto \exp( - j/\lambda)$:
\begin{equation}
\hspace{-1.5cm}\langle N\rangle C_j = \left\{ a_1 \left[ 1   - \Bigg| 1 -
2\left( \frac{j + \delta}{\omega}- Int\left( \frac{j + \delta}{\omega}
\right)\right)\Bigg| \right] - a_2 \right\}\cdot \exp \left( - \frac{j + \delta}{\lambda}
\right). \label{data}
\end{equation}
(where the parameter $\omega$ describes the observed periodicity period).  As shown in Fig. \ref{CMS_pseudorapidity} the amplitude of the oscillations is large for wide pseudorapidity windows and decreases with narrowing $|\eta|$. Also, the periodicity decreases with narrowing $|\eta|$; as a result, for small $|\eta|$ the oscillations vanish.

Fig. \ref{NBD_fc} shows that such oscillations are also present in the ALICE data \cite{ALICE1}. In this case the amplitude of the oscillations is even larger than in  the case of the CMS data, cf. Fig \ref{ALICE_CMS}. This is probably caused by the lower value of the $P(0)$ bin for the ALICE data (where $P(0) = 0.059$) than in the CMS (where $P(0) = 0.075$).

\begin{figure}[t]
\begin{center}
\resizebox{0.6\textwidth}{!}{%
 \includegraphics{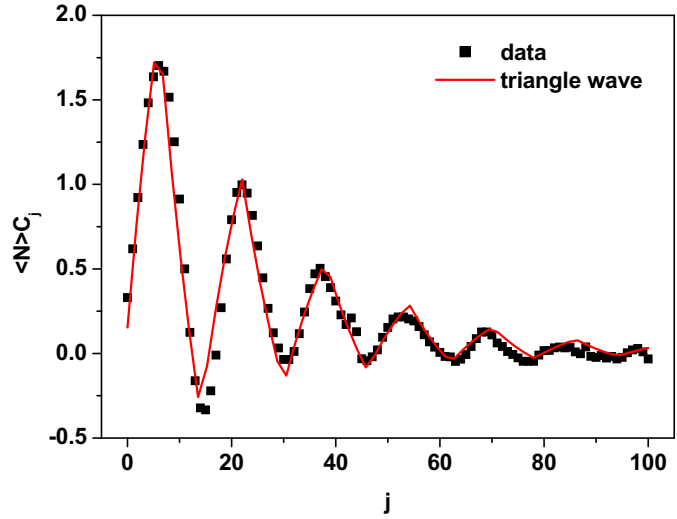}}
\vspace{-0.5cm}
\caption{(Color online) Example of the oscillations of coefficients $C_j$ describing the CMS data for $\sqrt{s} = 7$ TeV and pseudorapidity window $|\eta| < 2$ \cite{CMS}. They are fitted using the parametrization of $C_j$ given by Eq. (\ref{data}) with parameters $a_1 = 3.2$, $a_2 = 0.6$, $\omega = 16$, $\delta = 1.67$ and $\lambda = 25$.} \label{NBD_fa}
\end{center}
\end{figure}
\begin{figure}
\begin{center}
\resizebox{0.6\textwidth}{!}{%
 \includegraphics{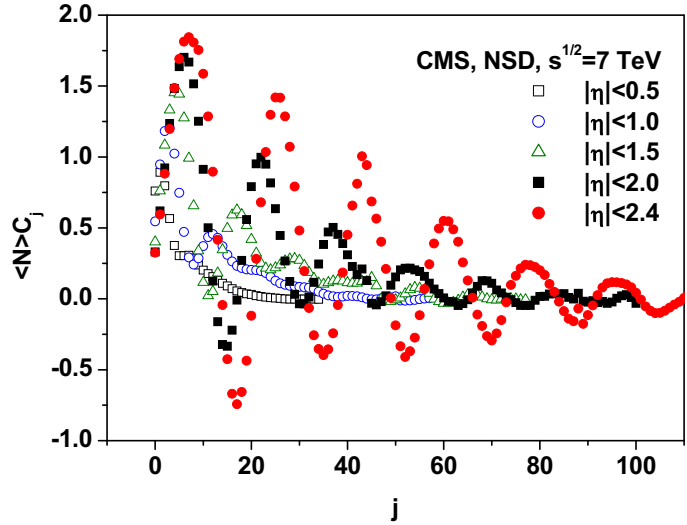}}
\vspace{-0.5cm}
\caption{(Color online) Coefficients $C_j$ emerging from the CMS data at $\sqrt{s} = 7$ TeV for different pseudorapidity windows \cite{CMS}. } \label{CMS_pseudorapidity}
\end{center}
\end{figure}
\begin{figure}[t]
\begin{center}
\resizebox{0.6\textwidth}{!}{%
\includegraphics{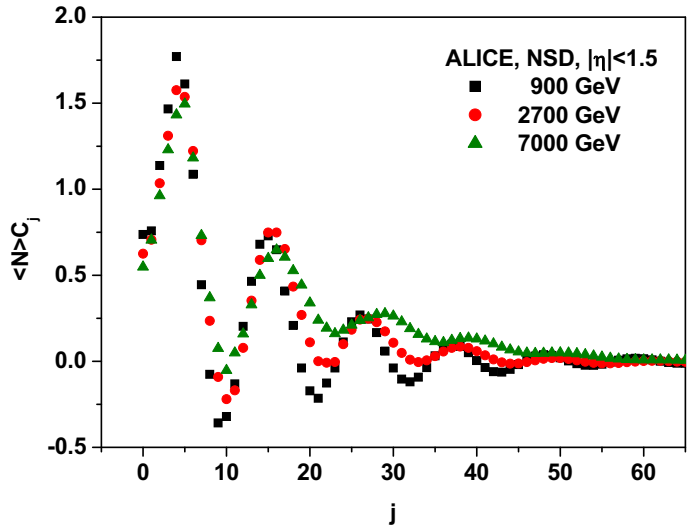} }
\vspace{-0.31cm}
\caption{(Color online) Coefficients $C_j$ emerging from the ALICE data \cite{ALICE1}.} \label{NBD_fc}
\end{center}
\end{figure}
\begin{figure}
\begin{center}
\resizebox{0.6\textwidth}{!}{%
\includegraphics{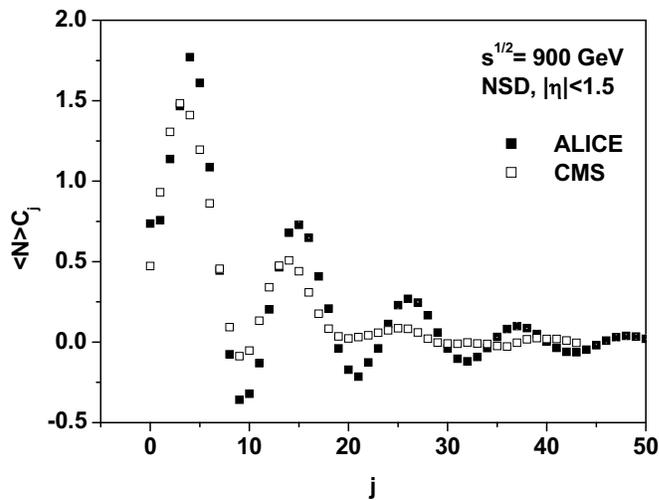} }
\vspace{-0.31cm}
\caption{(Color online) Comparison of coefficients $C_j$ emerging from the ALICE \cite{ALICE1} and CMS \cite{CMS} data taken for $\sqrt{s} = 900$ GeV and for $|\eta| < 1.5$ pseudorapidity window.} \label{ALICE_CMS}
\end{center}
\end{figure}

\begin{figure}[t]
\begin{center}
\resizebox{0.6\textwidth}{!}{%
\includegraphics{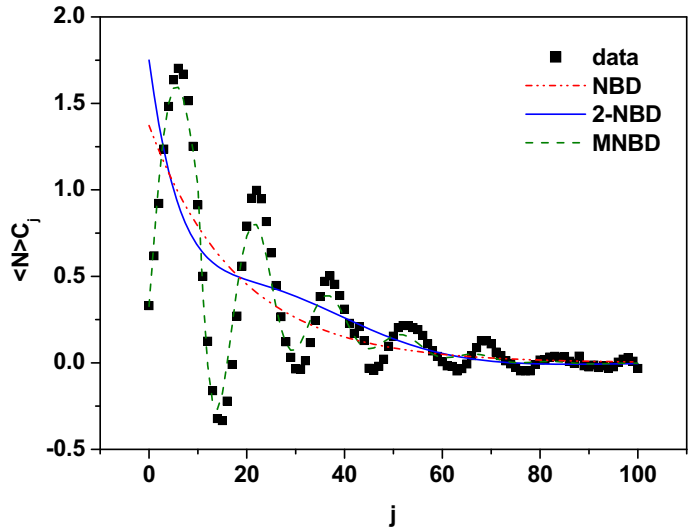} }
\vspace{-0.32cm}
\caption{(Color online) Coefficients $C_j$ emerging from the MNBD fit to the CMS data \cite{CMS} taken for $\sqrt{s}=7$ TeV and pseudorapidity window $|\eta|<2$ compared with the $C_j$ obtained from the single NBD and from the $2$-component NBD ($2$-NBD) fits to the CMS data with parameters from \cite{PG}.} \label{NBD_f}
\end{center}
\end{figure}
\begin{figure}
\begin{center}
\resizebox{0.6\textwidth}{!}{%
\includegraphics{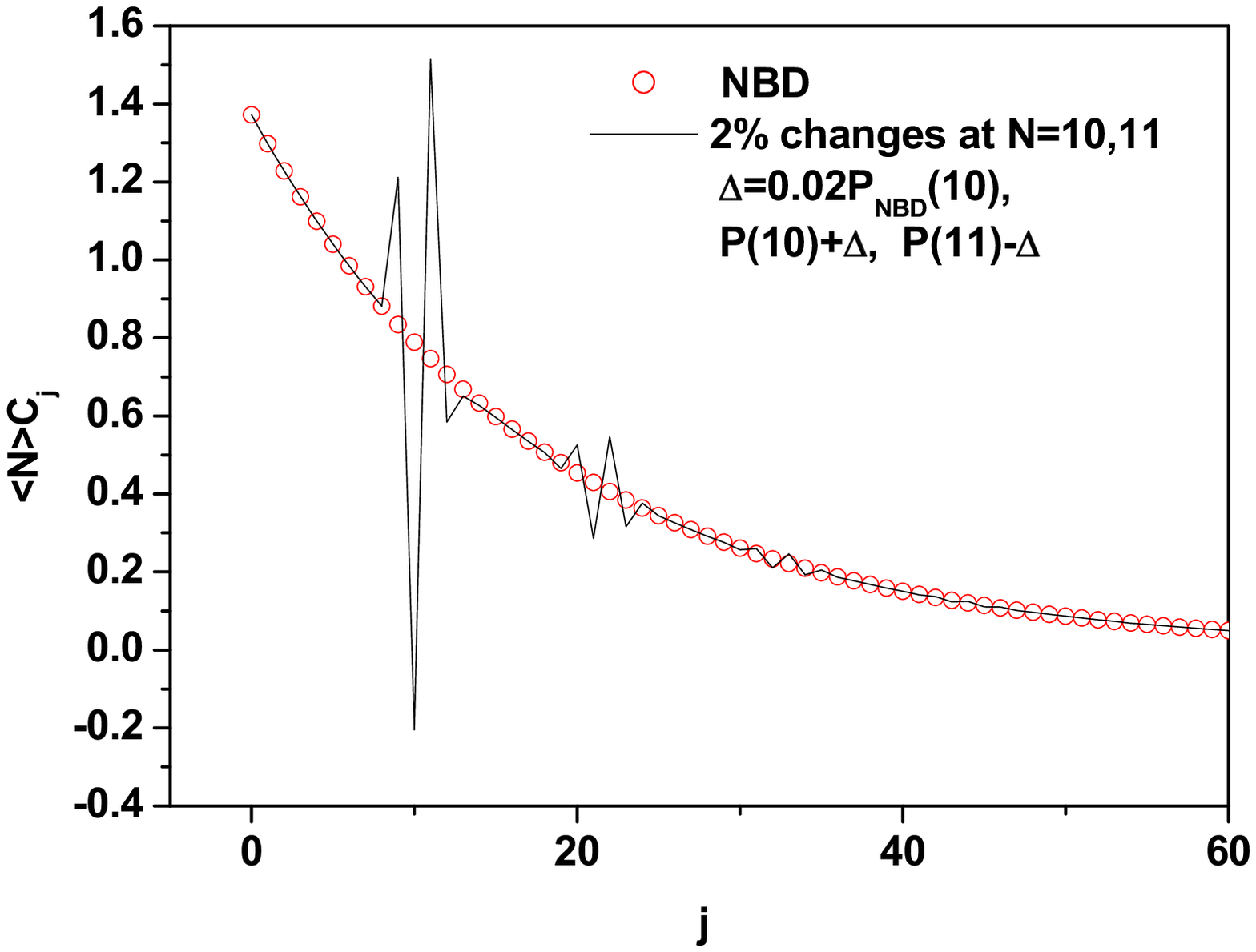} }
\vspace{-0.32cm}
\caption{(Color online) Illustration of how the oscillatory behavior of the coefficients $C_j$ emerges. See text for details.} \label{NBD_fd}
\end{center}
\end{figure}

However, as can be seen in Fig. \ref{NBD_f}, the corresponding coefficients $C_j$ calculated from the combination of the NBD distributions fitting the same data do not show such behavior. For single NBD they drop monotonically. This is because for the NBD given by Eq. (\ref{NBD}) the corresponding coefficients $C_j$ are equal to
\begin{equation}
C_j = \frac{k}{\langle N\rangle} p^{j+1} = \frac{k}{k+m}\exp(j\ln p),
\label{C_j_all}
\end{equation}
where $p$ is the probability of particle emission (\ref{partem}), which in this case is independent of the multiplicity $N$ and rank $j$~ \footnote{It is worth mentioning that such a form of $C_j$ when used in the recurrence relation (\ref{recPN}) reproduces relation (\ref{RR}) with $g(N) = \frac{mk}{m+k}\left(1 + \frac{N}{k}\right)$.}. The only dependence on the rank $j$ in Eq. (\ref{C_j_all}) is through the $p^j = \exp (j\ln p)$. Because $p <1$ always, the $C_j$ monotonically decrease, reproducing from the above fit (\ref{data}) only the damping exponent with the identification $\lambda = - 1/\ln p$~\footnote{It is interesting to note that in the case of a Binomial Distribution (BD) the corresponding coefficients are $C_j = (-1)^j\frac{k}{m}\left(\frac{m}{k - m}\right)^{(j+1)}$,
i.e., they oscillate very rapidly with a period equal to $2$. Note also that these $C_j$ do not depend on the multiplicity $N$.}.

The situation changes dramatically when we use probabilities of particle emission $p$ provided by the MNBD with parameters chosen to fit data \cite{CMS} and shown in Fig. \ref{Fig_PPE}. In this case the coefficients $C_j$ follow exactly the oscillatory behavior of the $C_j$ obtained directly from the CMS data, cf. Fig. \ref{NBD_f}.

Notice that coefficients $C_j$ evaluated from Eq. (\ref{Cj}) depend on $P(0)$.  In the experimental data the observed zeroth bin, $P(0)$, is very large (cf., for example, \cite{ALICE1}). Frequently we observe that $P(0) > P(1)$. On the other hand, there are no models showing that $P(0) > P(1)$ (for example, in the NBD one has that $P(0)/P(1) = 1/m +1/k  <1$ for reasonable choices of $k > m/(m-1)$). This fact is not much discussed. In fact, because of the experimental difficulties, this bin is frequently omitted in the analysis of data \cite{ALICE1}. On the other hand it should be remembered that $P(0)$ is the only bin which is very sensitive to the acceptance (cf. \ref{app}). In our case, for the proper normalization we used for $P(0)$ in Eq. (\ref{Cj}) for MNBD shown in Fig. \ref{NBD_f} value $P(0) = 0.045$, which corresponds to the acceptance probability $\alpha = 0.965$.

Fig. \ref{NBD_fd} shows how such oscillations can occur. In the example shown there we have used the NBD (which, as was shown before, itself results in exponentially decreasing $C_j$) in which we have made the following changes: we put $P(10) = P_{NBD}(10) + \Delta $ and $P(11) = P_{NBD}(11) - \Delta $ with $\Delta = 0.02 P_{NBD}$. As seen in Fig. \ref{NBD_fd} these small and apparently innocent changes resulted in rather dramatic spikes occurring on the original $P_{NBD}$ (circles) with period $\sim 10$ and with rapidly falling sizes. This means that the coefficients $C_j$ are very sensitive to all changes in the original $P(N)$. Precisely such a change is provided by the MNBD, for which the probability of emission $p$ is no longer constant but shows a characteristic spike clearly visible in Fig. \ref{Fig_PPE}. This spike influences then, via Eq. (\ref{Cj}), the consecutive coefficients $C_j$ and brings them to agreement with those obtained from the experimentally measured $P(N)$.

Note that the coefficients $C_j$ tell us how $P(N+1)$ depends on $P(N-j)$, i.e., they encode the memory about particles produced earlier. In the case of the NBD, as given in Eq. (\ref{C_j_all}), this memory exponentially disappears with increasing distance (rank) $j$. Fig. \ref{NBD_fd} shows that in the example considered there this loss of memory is non-monotonic. The falloff is exponential but it is now decorated with characteristic oscillations. We still feel particles located $j = - 8 + 15.4 i$ away from the multiplicity in question ($i$ denotes  the consecutive maxima). This means that for some multiplicities (or, for some distances from the multiplicity of interest to us) the encoded memory is stronger. Such behavior strongly indicates that particles are produced in clusters. The fact that for large $j$ one has smaller $C_j$ means only that the further away we are from a given multiplicity the smaller is its influence on the final distribution.

\begin{figure}[t]
\begin{center}
\resizebox{0.6\textwidth}{!}{%
\includegraphics{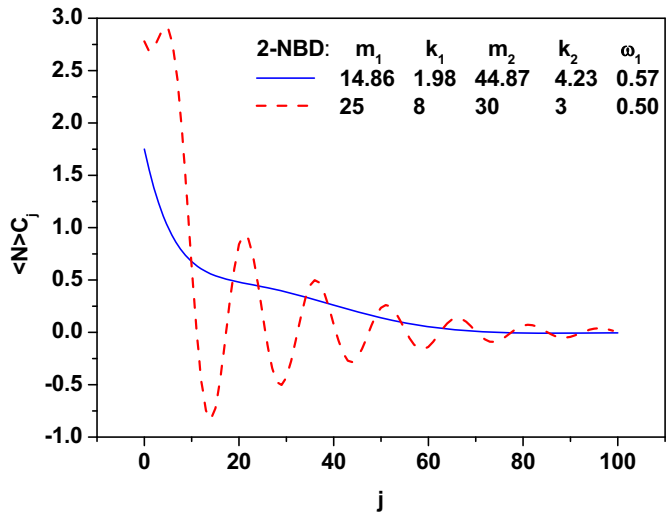} }
\vspace{-0.32cm}
\caption{(Color online) Coefficients $C_j$  emerging from the $2$-component NBD with parameters from \cite{PG} (solid line) and with some arbitrarily chosen parameters leading to oscillations (dashed line).} \label{MNBD_f}
\end{center}
\end{figure}

 We close this section by noting that, after all, multicomponent NBD distributions can lead to oscillatory behavior of the coefficients $C_j$ in some special circumstances. Let
\begin{equation}
P(N) = \sum_i \omega_i P_{NBD}\left(N,p_i\right) \label{MNBD}
\end{equation}
be a superposition of a number of NBD with weights $\omega_i$ (such that $\sum_i \omega_i = 1$) and emission probabilities $p_i = m_i/\left(m_i + k_i\right)$. In this case Eq. (\ref{Cj}) should be replaced by the following recurrence relation:
\begin{eqnarray}
\hspace{-2.5cm} P(0) C_j &=& \frac{(j+1)}{\langle N\rangle} \sum_i \omega_i P_{NBD}\left(j+1,p_i\right) - \sum_{l=0}^{j-1}C_l \sum_i\omega_i P_{NBD}\left(j-l,p_i\right)=\nonumber\\
\hspace{-2.5cm} &=& (j+1)\left[ \sum_i \omega_i P_{NBD}\left(j+1,p_i\right)\left( \frac{1}{\langle N\rangle} - \frac{1}{m_i}\right)\right] + \sum_i \omega_i \left( 1 - p_i\right)^{k_i + 1} p_i^j + \nonumber\\
\hspace{-2.5cm} && + \sum_i\omega_i\sum_{l=0}^{j-1}\left[\left(1 - p_i\right) p_i^l - C_l\right]
P_{NBD}\left(j-l,p_i\right). \label{MNBD1}
\end{eqnarray}
Using now Eq. (\ref{NBD}) for $P_{NBD}\left(N,p_i\right)$ one gets that
\begin{eqnarray}
\hspace{-2.5cm} C_j &=& \frac{1}{P(0)} \sum_i \omega_i p_i^j \left(1 - p_i\right)^{k_i +1}\cdot \nonumber\\
\hspace{-2.5cm} &\cdot& \left\{ \frac{\Gamma\left(j+k_i+1\right)}{\Gamma\left( k_i\! +\! 1\right)\Gamma(j+1)}\frac{m_i\! -\! \langle N\rangle}{\langle N\rangle}\! +\! 1\! +\! \sum_{l=0}^{j-1}\left[ 1\! -\! \frac{C_l}{p_i^l\left( 1 - p_i\right)}\right] \frac{\Gamma\left(j\! -\! l\! +\! k_i\right)}{\Gamma\left(k_i\right)\Gamma\left(j - l +1\right)} \right\}. \label{MNBD2}
\end{eqnarray}
Note that for  $m_i < \langle N\rangle$, as well as for $C_l > p_i^l\left(1 - p_i\right)$, we have negative terms which can result in nonmonotonic behavior of the coefficients $C_j$. An example of such behavior is shown in Fig. \ref{MNBD_f}}. We compare there the $C_j$ obtained from the $2$-component NBD used to describe experimental data (and claimed to do so successfully) \cite{PG} with the $2$-component NBD with parameters chosen in such a way as to obtain oscillatory behavior of $C_j$ but without attempting to fit experimental data. One can therefore summarize this point by stressing that possible successful models of multiparticle production  should describe, with the same parameters, both the multiplicity distributions, $P(N)$, and the corresponding coefficients, $C_j$, because these coefficients provide us with new information which can be used to improve models  of particle production processes.

\section{Summary and conclusions}
\label{SaC}

The multiplicity distributions $P(N)$ provide an indispensable tool in the investigation of the dynamics of multiparticle production processes. Their measurement forms an important part of the experimental activity. Likewise, the extraction of useful information from these measurements, its proper identification and understanding, is one of the main objects of theoretical studies. So far, despite the abundance of models of $P(N)$ (more or less successful), the problem of the description of the observed properties of $P(N)$ is still unsolved. We do not have either a convincing physical interpretation of the origin of the large fluctuations of $P(N)$ or a fine structure of the charged particle statistics detected in inelastic collisions. In this paper we argue that all experimental multiplicity distributions of secondary particles measured in hadron interactions and  the fine structure of these distributions detected experimentally  can be analysed in terms of a suitable recurrence relation. In particular, we argue that it is possible to retrieve from the measured $P(N)$ some additional information on the dynamics of the multiparticle production processes to that already investigated. Our motivation was the unprecedented popularity of the NBD as a tool describing the measured multiplicity distributions, essentially in all reactions and at all energies. However, this success seems to be endangered because of the small, but noticeable and rather systematic discrepancies (which apparently grow with energy). To address them one has to study the effects of possible modifications which could improve agreement with data.

At first, it turned out that to get a flat ratio $R = P_{CMS}(N)/P_{MNBD}(N)$ (see Fig. \ref{Fig_Modified}) one has to introduce a rather striking, non-monotonic dependence of the parameter $m$ on the multiplicity $N$ (Eq. (\ref{Non}) ) resulting in a very pronounced non-monoticity of the probability of emission ratio $p$ as function of $N$, shown in Fig. \ref{Fig_PPE}. For the time being this must be left as a phenomenological invention without any theoretical justification to hand.

In the second part our tool is the recurrence relation (\ref{recPN}). However, it must be stressed that we are not attempting to describe the measured multiplicity distribution, $P(N)$, by first calculating  the coefficients $C_j$ in Eq. (\ref{recPN}) from experimental data on, for example, rapidity distributions (as was the case in \cite{CSF,CSF1}). Instead, we use the observed $P(N)$ as our input and calculate the corresponding coefficients $C_j$ by means of Eq. (\ref{Cj})\footnote{It must be noted at this point that we are tacitly assuming that the measured $P(N)$ are fully reliable. The possible sensitivity of the coefficients $C_j$  to the systematic uncertainties of the measurement and to the unfolding uncertainties of the experimental procedure used, can be checked only by the careful analysis of the raw data, using the proper response matrix, and that exceeds our capability.}. As a result, we have discovered that the coefficients $C_j$ are very sensitive to the details of the shape of the multiplicity distribution $P(N)$ used, here to the details of the MNBD. We argue therefore that detailed analysis of the coefficients $C_j$ will provide additional valuable information to that which can be obtained by the usual fitting of the $P(N)$ alone. The best example is the fact that, so far, the $2-$NBD which fits data is not able to describe the coefficients $C_j$ obtained from the experimental $P(N)$ (although, as was demonstrated above, in principle oscillations of $C_j$ are possible in a multi-component NBD scenario). Another point is that, as shown in \ref{app}, $g(N)$ defining the MD in the recurrence relation (\ref{RR}) does not depend (for $N > 0$) on the acceptance. This is the reason why the approach using it is attractive for the description of experimental data which have a constant problem with measuring $P(0)$.

So far, the only other similar results (known to us) were found in works using  combinants, $C^{\star}_j$, introduced in \cite{KG} and defined for a generating function $G(z)$ as
\begin{equation}
C^{\star}_j = \frac{1}{j!} \frac{d^j \ln G(z)}{dz^j}\Bigg|_{z=0}. \label{combinants}
\end{equation}
They were first used in \cite{JB} and found to be useful for a detailed study of the low-multiplicity part of the distribution and for the description of the multiplicity distributions of rare particles. In \cite{BS} they were shown to be a useful tool in identifying  the nature of the source of the emitted particles whereas in \cite{Bao} they were applied to pion multiplicity distributions in heavy ion collisions at low energies.  Finally, in \cite{Hegyi1,Hegyi2,Hegyi3} they were used to analyse some modified versions of the NBD (which, as required by the higher-order perturbative QCD effects,  violated their usual infinite divisibility property).

It turns out that our coefficients $C_j$ are directly connected to the combinants $C^{\star}_j$ (cf. \ref{CjC}), namely
\begin{equation}
C_j = \frac{(j+1)}{\langle N\rangle}C^{\star}_{j+1}. \label{CC}
\end{equation}
Note that the $C_i$ are much more sensitive to oscillations than combinants (especially for higher values of index $j$). Both $C_j$ and $C_j^{\star}$ can be also applied to correlations, namely to the cumulants of the multiplicity distributions (for NBD these coefficients are simply related to the $H_q$ moments)\footnote{Usual factorial moments and cumulants do not describe data (cf. \cite{Sar,OPAL}) and the usual $H_q$ moments (ratios of cumulants to factorial moments) for NBD do not show any oscillations \cite{DG}.}.

To summarize: we propose a novel phenomenological description of the observed multiplicity distributions which allows for a more detailed quantitative description of the complex structure of the experimental data on $P(N)$. It is provided by the coefficients $C_j$ defining  the recurrence relation (\ref{recPN}). However, it should be noted that our analysis is not directly connected with the wave structure observed in data on $P(N)$ for multiplicities above $N=25$ \cite{ALICE1}. The coefficients $C_j$ are completely insensitive to the $P(N > (j+1))$ tail of the multiplicity distribution, while the oscillatory behavior of $C_j$ is observed starting from the very beginning. The wave structure of the multiplicity distributions already observed by ALICE, CMS (and previously also by UA5 \cite{UA5}) experiments is still hardly significant; in fact it is suspected to occur as an artifact of the unfolding procedure used in each experiment, with the period related to the corresponding width of the response matrix \cite{ALICE1}. Sensitivity of coefficients $C_j$ to the systematic uncertainties of the measurement and to the unfolding uncertainties can be checked only by detailed analysis of the raw data with the proper response matrix, which is beyond the scope of the present work. However, if these oscillations are be experimentally confirmed, they will await their physical justification, i.e., indication of some physical process which would result in such a phenomenon.

\section*{Acknowledgment}

We are indebted to Sandor Hegyi, Edward Grinbaum-Sarkisyan and Adam Jacho\l kowski for fruitful discussions and we would like to thank warmly Nicholas Keeley for reading the manuscript.
The research of GW was supported in part by the National Science Center (NCN) under contract Nr 2013 /08/M /ST2 /00598 (Polish agency).

\appendix

\section{The problem of $P(0)$ }\label{app}

The NBD can also be defined by the following probability generating function
\begin{equation}
G_{NBD}(z) = \left( \frac{1 - p}{1 - p z} \right)^k\quad{\rm
where}\quad p =\frac{m}{m+k}. \label{NBDGF}
\end{equation}
Particles are registered with probability $\alpha $ and their acceptance process is described by the binomial distribution with generating function
\begin{equation}
G_{BD}(z) = 1 - \alpha + \alpha z. \label{BDGF}
\end{equation}
The number $N$ of registered particles is
\begin{equation}
N = \sum^M_{i=1} n_i. \label{NRP}
\end{equation}
where $n_i$ follows the BD and $M$ comes from the NBD. The generating function for the distribution of $N$ registered particles is then given by
\begin{equation}
G(z) = G_{BD}\left( G_{NBD}(z)\right) = 1 - \alpha + \alpha
\left(\frac{1 - p}{1 - p z}\right)^k . \label{Gboth}
\end{equation}
This corresponds to a probability distribution of registered particles
\begin{equation}
P(N) = \frac{1}{N!} \frac{d^NG(z)}{d s^N}\Bigg|_{z=0}.
\label{P_sum}
\end{equation}
The corresponding recurrence relation for this distribution is
\begin{equation}
g(N) = \frac{(N+1)P(N+1)}{P(N)} = \frac{\frac{d^{N+1} G(z)}{d
z^{N+1}}\Big|_{z=0}}{\frac{d^N G(z)}{d z^N}\Big|_{z=0}}.
\label{Ratio}
\end{equation}
Note that for $N > 0$ the function $g(N)$ does not depend on the acceptance and is the same as that for the NBD. However, for $N=0$ the acceptance $\alpha$ enters and one has that
\begin{equation}
g(0) = \frac{\alpha \left( 1 - p\right)^k pk}{1 - \alpha + \alpha
\left( 1 - p\right)^k} = \frac{m\alpha \left( \frac{k}{m + k}\right)^{k+1}}{1 - \alpha + \alpha\left( \frac{k}{m + k}\right)^k}.\label{g(0)}
\end{equation}
In fact, the above result is valid for any distribution $P(M)$ with probability generating function $G_M(s)$, i.e., the term with $N=0$,
\begin{equation}
g(0) = \frac{\alpha }{1 - \alpha + \alpha G_M(0)}
\frac{dG_M(z)}{dz}\Bigg|_{z=0}, \label{general}
\end{equation}
always depends on the acceptance.

\section{Coefficients $C_j$ and combinants $C^{\star}_j$}\label{CjC}

We present for completeness a derivation of relation (\ref{CC}). To start with, let us note that whereas the $C_j$ are defined by the recurrence relation given by Eq. (\ref{recPN}), combinants are defined by the generating function $G(z)= \sum^{\infty}_{N=0}P(N)z^N$ \cite{KG},
\begin{equation}
\ln G(z) = \ln P(0) + \sum_{j=1}^{\infty} C^{\star}_j z^j \label{KG}
\end{equation}
(and are regarded as a measure of departure from the pure Poissonian distribution \cite{KG, VVP}). This means, therefore, that in terms of $C^{\star}_j$
\begin{equation}
G(z) = P(0)\exp\left[\sum^{\infty}_{j=1}C^{\star}_jz^j\right]\quad{\rm and}\quad P(N) = \frac{1}{N!}\frac{d^NG(z)}{dz^N}\bigg|_{z=0} \label{GPCstar}
\end{equation}
resulting in the recurrence relation
\begin{equation}
(N+1)P(N+1) = \sum^{N}_{j=0}(j+1)C^{\star}_{j+1} P(N-j). \label{recstar}
\end{equation}
Comparing Eq. ({\ref{recstar}) with Eq. (\ref{recPN}) we get that
\begin{equation}
\langle N\rangle C_j = (j+1)C^{\star}_{j+1}, \label{CstarCfin}
\end{equation}
i.e., Eq. (\ref{CC}).

\end{document}